\documentstyle{article}[12pt]

\def\be{\begin{equation}}
\def\ee{\end{equation}}
\def\ben{\begin{displaymath}}
\def\een{\end{displaymath}}
\def\ba{\begin{array}{c}}
\def\ea{\end{array}}

\begin{document}


\vspace*{1cm}

\begin{center}
{\Large\bf Bound states in the Kratzer plus polynomial potentials
and their new exact tractability via nonlinear algebraic equations
}
\end{center}


\begin{center}
Miloslav Znojil


OTF, \'{U}stav jadern\'e fyziky AV \v{C}R, 250 68 \v{R}e\v{z},
Czech Republic

\end{center}

\vspace{5mm}



\section*{Abstract}

Schr\"{o}dinger equation with potentials of the Kratzer plus
polynomial type (say, quartic $V(r) = A\,r^4 +B\,r^3 + C\,r^2+D\,r
+F/r +G/r^2$ etc) is considered and a new method of exact
construction of some of its bound states is presented. Our
approach is made feasible via a combination of the traditional use
of the infinite series $\psi(r)$ (terminated {\em rigorously}
after $N+1$ terms at certain specific couplings and energies) with
several new ideas. We proceed in two steps. Firstly, in the
strong-coupling regime with $G \to \infty$, we find the exact,
complete and compact unperturbed solution of our $N+2$ coupled and
{\em nonlinear} algebraic conditions of the termination. Secondly,
we adapt the current Rayleigh-Schr\"{o}dinger perturbation theory
to our nonlinear equations and define the general $G < \infty$
bound states via an innovated, {\em triple} perturbation series.
In its tests we show how all the corrections appear in integer
arithmetics and remain, therefore, exact.

 \vspace*{1cm}

\newpage

\section*{I. Introduction}

Bound states in the Coulomb or, after a slight generalization,
Kratzer's  \cite{Kratzer} potential
 \ben
  V_{\{0\}}(r)= F/r+G/r^2
 \een
played an important role in the history of quantum mechanics
\cite{tenpan} as well as in the various methodical studies of its
limitations \cite{Frank}.  In applications the potential offered
one of the most important exactly solvable models of atomic and
molecular physics and quantum chemistry. The model is extremely
transparent, its spectrum of energies $E$ is numbered by integers
$N = 0, 1,\ldots$ and its wave functions are proportional to
Laguerre polynomials of degree~$N$~\cite{Fluegge}. Almost 30 years
ago A. Hautot noticed \cite{Hautot} that the exact solvability of
the Kratzer's potential survives, in a way, its immersion in the
shifted harmonic confining well,
 \ben V_{\{1\}}(r)= C\,r^2+D\,r+F/r+G/r^2.
 \een
For {\em any integer} $N = 0, 1,\ldots$ one may construct $N+1$
different ``Sturmian" exact wave functions proportional to
polynomials of degree $N$.  In contrast to the Kratzer's model the
Hautot's elementary bound states do not form a complete set. In
literature the solutions of this type are called quasi-exact
\cite{Ushveridze}.

The two classes of forces $V_{\{0\}}(r)$ and $V_{\{1\}}(r)$ are
usually considered exceptional. This attitude finds a deeper
mathematical foundation in the representation theory of Lie
algebras \cite{Turbiner}. Still, certain remnants of their
elementary solvability may be detected in all their further
polynomial extensions. This has been first noticed by Magyari
\cite{Magyari} who generalized the Hautot's method to all the
polynomial forces
 \be
V_{\{m\}}(r)= A\,r^{2m} + B\,r^{2m-1} + \ldots + F/r+G/r^2.
\label{genpot}
 \ee
The generalized construction with degree $m\geq 2$ did not
inspire, unfortunately, too many applications. Basically, this was
due to its numerous practical shortcomings. In introduction, let
us only mention that Magyari replaced the {\em linear}
Schr\"{o}dinger equation by a {\em nonlinear} algebraic system.
The solution of these equations seems only feasible for the first
two choices of indices $m=0$ and $m=1$, with nonlinearity more or
less kept under control.  In the present paper we intend to
demonstrate that such a constraint and {\it a priori} scepticism
about the feasibility of solution of the Magyari's equations are
undeserved, for the important \cite{quartici} choice of $m=2$ at
least.

The paper is inspired by certain unpublished numerical experiments
with Magyari equations (\cite{Dubna}, cf. also Section II). In the
first step of a new development we show in Section III that
certain unsolvable interactions (\ref{genpot}) with $m=2$ and with
a strong repulsive core $G\to \infty$ {\em leave the Magyari's
equations exactly solvable at any dimension} $N$. This is our key
result.

In the equally important second step we demonstrate how one can
deal with the corrections for finite $G < \infty$. Section IV
offers a new perturbation method of construction of their sequence
in closed form. The method emphasizes several close parallels
between our nonlinear algebraic problem and its simpler linear
analogues. Technical details of our Rayleigh-Schr\"{o}dinger-like
new perturbation theory are illustrated on our quartic example
$V_{\{2\}}(r)$ in full detail. Section V adds a few concluding
remarks.

\section*{II. Matrix form of the Schr\"{o}dinger equation \label{I}
\label{I.1} \label{I.2}}

An overall methodical background of the Magyari's treatment of the
ordinary radial Schr\"{o}dinger equation
\be
\left[-\,\frac{d^2}{dr^2} + \frac{\ell(\ell+1)}{r^2} + V(r)
 \right]\, \psi(r) =
E \psi(r)
 \label{rad}
 \ee
with angular momentum $\ell=0, 1, \ldots$ lies somewhere in
between the variational and non-variational approaches. While one
usually expands wave functions in an appropriate orthonormalized
basis, Magyari uses the mere Taylor series.  His recipe weakens
the current emphasis upon methodical universality and numerical
efficiency. It favors, instead, an enhanced simplicity,
analyticity and closed, non-numerical form of his quasi-exact
bound states in a way which is re-gaining a new credit in recent
literature \cite{Brihaye}.

At the very beginning of our considerations let us note that all
the asymptotically polynomial forces (\ref{genpot}) with $ m \geq
1$ imply the same asymptotically exponential form of their
bound-state wave functions,
 \ben
 \psi(r)
=\exp \left\{-\sqrt{A}\,\left [r^{m+1}/(m+1) +B\,r^m/2A +{\cal O}
\left (r^{m-1} \right ) \right ] \right \}.
 \een
This is an important information. Once we restrict our attention
to the mere quartic $m=2$ potential for the sake of definitness
and brevity,
 \be V_{\{2\}}(r)= A\,r^4 + B\,r^3+  C\,r^2 + D\,r+
F\,r^{-1}+ G\,r^{-2}
\label{quartic}
 \ee
we may try to write down the polynomial ansatz
\be
\psi_{\{2\}}(r)=\exp \left( -\, \frac{1}{3} \alpha_{} \,r^3 -
 \frac{1}{2} \beta \,r^2 - \gamma \,r \right)\
\sum^{N}_{n=0} \omega_n\,r^{n+l+1}.
 \label{anhari}
 \ee
At an arbitrary finite integer $N \geq 0$ the requirement of its
asymptotically correct behaviour fixes the parameters $\alpha
=\sqrt{A} > 0$, $\beta_{}=B /2\alpha_{}$ and $\gamma_{}=(C
-\beta_{}^2) /2\alpha_{}$.  The next order of asympotic analysis
also reveals the uniqueness of the $N-$dependent
``termination-admitting" value of the coupling constant $D =
D(N)=-2\alpha(N+l+2)+2\beta \gamma$. This parallels the Hautot's
observations made at~$m=1$~\cite{Hautot}.

Near the origin the ansatz (\ref{anhari}) combines the angular
momentum $\ell = 0, 1, \ldots$ with the centrifugal-like coupling
$G>-1/4$ into a new quantity $l=l(G)$ such that $G +\ell(\ell+1) =
l(l+1)$. After the removal of a sign ambiguity in $l=-1/2
+\sqrt{G+(\ell+1/2)^2}>-1/2$ we may insert our ansatz in eq.
(\ref{rad}) with potential (\ref{quartic}). This transforms the
differential Schr\"{o}dinger equation into equivalent recurrences
for Taylor coefficients $\omega_j$,
 \be
R_k\omega_{k-2}+ T_k\omega_{k-1}+
 S_k\omega_{k}+
 P_k\omega_{k+1}
 = 0  \label{4.8}
 \ee
with $ k = 0, 1, \ldots, N+1$ and coefficients
 \ben
  R_n = 2 \alpha (
N+2-n), \ \ \ \ \ \ T_m=E+\gamma^2-\beta(2m+2l+1)
 \een
(abbreviated as $T_m \equiv 2\,T-2(m-1)\beta$) and
 \ben
  S_m=-
2\gamma(m+l+1)-F , \ \ \ \ \ \
 P_m=
(m+1) (m+2l + 2)
 \een
(with abbreviations $S_m=\equiv 2\,S-2\gamma\,m$ and $l+1
\equiv\Omega $). These equations form an over-determined linear
system of equations for parameters $\omega_j$ with asymmetric and
non-square four-diagonal matrix $N+2$ by $N+1$,
 \ben \left(
\begin{array}{ccccc}
S&\Omega&&&
 \\  &&&&\\T&S-\gamma&2\Omega+1&&
  \\ &&&&\\
N \alpha &T-\beta&S-2\gamma&3\Omega+3&
 \\
 &&&&\\
 & (N-1)\alpha
 &T-2\beta  &
 \ddots\ \ \ \ \ \ \ \ \ \ \ \ \ \ \ \ \ \
  &\ddots \ \ \ \ \ \ \ \ \ \ \ \ \ \ \ \ \ \ \\
  & \ \ \ \ \ \ \ \ \ \ \
 \ddots& \ \ \ \ \ \ \ \ \ \ \
 \ddots&S-(N-1)\gamma& N \Omega +  \left ( \ba
  N\\2 \ea \right )
   \\ & & 2\alpha& T-(N-1)\beta&
  S-N\gamma
 \\
 &&&&\\ &
& & \alpha& T-N\beta\\
\end{array} \right). \label{eq4.8}
 \een
Within the approach based on the latter equations one only works
with a few exceptional bound states. They are elementary and
preserve a strong formal similarity to the popular harmonic
oscillators. Unfortunately, this parallel seems to weaken with the
growth of $N$ since the explicit solution of our non-linear
problem (\ref{4.8}) quickly becomes more and more complicated.

We intend to simplify the equations in question at a cost of their
perturbative re-arrangement. Such an attempt is motivated by the
possibility of a suppression of the most quickly growing
(binomial) matrix elements at large $\Omega\equiv l(G)+1\sim
G^{1/2}\gg 1$, i.e., under the presence of a strong repulsive core
in potential (\ref{quartic}). A routine re-scaling of coordinates
$ r \to \mu\,r$ in eq. (\ref{rad}) and of the related coefficients
$\omega_n \to u_n=\omega_n \mu^n$ in eq. (\ref{anhari}) leads to a
modified and more transparent matrix form $Q\,\vec{u}=0$ of the
Magyari's eq.~(\ref{4.8}),
 \be
 \left(
\begin{array}{cccc}
S&\Omega/\mu&&  \\ T\mu&S-\gamma&2\Omega/\mu+1/\mu &
\\ \alpha N\mu^2&(T-\beta)\mu&
S-2\gamma&\ddots\ \ \ \ \  \ \
\\
  & &  &
\ddots\ \ \ \ \ \ \  \\ &\ \ \ \ \ \ \ \  \ \ \ \ddots&\ \ \ \ \ \
\ \ \ \ \ \ddots&\\
 &  &  \alpha\mu^2& (T-N\beta)\mu\\
\end{array} \right)
\left ( \ba u_0\\u_1\\ \vdots\\ u_N
 \ea \right ) = 0 .
 \label{scaled}
 \ee
In the repulsive $\Omega\gg N$ regime we notice that $|S| \gg
N\,|\gamma|$ and $|T| \gg  N\,|\beta|$. The two main diagonals of
$Q$ become approximately constant.

With the energy $E$ proportional to the new parameter $T=T(E)$ and
with the Coulomb coupling $F$ contained in $S=S(F)$ a ``symmetry"
of $Q$ may be enhanced by the choice of the value of the scaling
parameter $\mu$ in such a way that the magnitude of its uppermost
and lowest diagonals is balanced.  For example, with
$\alpha\mu^2=\Omega/\mu= \tau$ we fix the unique values of $\mu =
(\Omega/\alpha)^{1/3}$ and $\tau = (\Omega^2\alpha)^{1/3}$.  Then
we may abbreviate $s =S/\tau$ and $t=\mu T/\tau$ and pre-multiply
and split our pseudo-Hamiltonian into a sum of matrices $ Q =
Q^{(0)} +\lambda\,Q^{(1)}$. Its first component
 \ben
 Q^{(0)}= Q^{(0)}(s,t)=
 \left(
\begin{array}{cccccc}
s&1&&& & \\ t&s&2 && & \\ N&t&s&3 &&
\\ &\ddots&\ddots&\ddots &\ddots& \\ & &3 & t&s &N\\ && & 2&t&s\\
& & & & 1&t\\
\end{array} \right)
 \een
will play the role of our unperturbed Magyari-Schr\"{o}dinger
pseudo-Hamiltonian. In the strong-core limit $G \to \infty$ our
zero-order Magyari equation reads
 \be
  Q^{(0)}(s,t)
\left( \ba u^{(0)}_0\\ u^{(0)}_1\\  \vdots\\ u^{(0)}_N \ea \right)
= 0.
 \label{finale}
 \ee
It has the form of an over-complete system of $N+2$ equations
which should determine all the $N+1$ Taylor coefficients
$u^{(0)}_n$ plus the two spectral-like parameters $s=s^{(0)}$ and
$t=t^{(0)}$. In the purely mathematical setting, the difficulties
represented by non-square form of their pseudo-Hamiltonian $
Q^{(0)}(s,t)$ are quite new and rarely encountered in the mainly
linear formalism of the traditional quantum mechanics.  Of course,
the solvability of eq. (\ref{finale}) with $m \geq 2$ loses a
natural Lie-algebraic background and interpretation of its
quasi-exact predecessors with $m \leq 1$ \cite{Turbiner}.


\section*{III. Unperturbed solutions
 \label{II}}

At an arbitrary $N$, let us introduce the two non-square
quasi-unit matrices ${\cal J}$ and ${\cal K}$ such that
$Q^{(0)}(s,t)= Q^{(0)}(0,0)+s\,{\cal J} + t\,{\cal K}$. Denoting
their transposition by a superscript $^T$ it is easy to imagine
that the products $ {\cal J}^T\,Q^{(0)}(s,t)$ and ${\cal
K}^T\,Q^{(0)}(s,t)$ are square matrices.  In such a notation the
necessary and sufficient condition of the nontrivial solvability
of eq. (\ref{finale}) may be formulated as a simultaneous
disappearance of the two independent secular determinants, say,
\be
\det \left [{\cal J}^T\,Q^{(0)}(0,t)+s\,I\right ]=0, \ \ \ \ \ \ \
\det \left [{\cal K}^T\,Q^{(0)}(s,0)+t\,I\right ]=0.
 \label{finalab}
  \ee
Our two free unperturbed parameters $s=s^{(0)}$ and $t=t^{(0)}$
will be fixed and determined by these two coupled polynomial
equations at any $N$ in principle.

Due to symmetries of our generalized, coupled eigenvalue problem
(\ref{finalab}) the pairs of its (in general, complex) solutions
will always remain complex conjugate, $s^{(0)} =\left
[t^{(0)}\right ]^*$. For physical reasons we must ask whether at
least some of these solutions remain real, $s^{(0)}=t^{(0)}$. The
answer may be found by their explicit evaluation.

At $N=0$ the solution of eq. (\ref{finalab}) is unique. From the
natural normalization $u^{(0)}_N=1$ we only get the trivial
$s=t=0$. At $N=1$ the same normalization implies that
$u^{(0)}_0=-1/s$ (e.g., via the first row of eq. (\ref{finale})).
The second row defines $t=s^2$ and our set degenerates to the
single cubic equation $s^3=1$. Out of its three different complex
roots only one is real and we have the unique physical solution
$s=t=1$. At the next dimension $N=2$ an alternative, intermediate
normalization $u^{(0)}_1=1$ helps us to eliminate $u^{(0)}_0=-1/s$
and $u^{(0)}_2=-1/t$ via the respective first and last row of eq.
(\ref{finale}). This gives $s^3=t^3$ and linear relation $t = a\,s
$ with the three possible complex constants $a$ such that $a^3=1$.
One ends up with a triplet of alternative quadratic equations for
the unknown $s$. Similarly, one proceeds at the higher
integers~$N$.

The resulting sets of the roots $s$ are sampled in Table
\ref{list}. The general pattern of their $N-$dependence is
obvious. Once we pay attention to the real solutions only, we
arrive at the general formula
 \be
s=t=s^{(0)}_{[n+1]}(N)=t^{(0)}_{[n+1]}(N)=N-3n,\ \ \ \ \ \ \ \
 n = 0,1, \ldots, [N/2].
 \label{fyzi}
 \ee
We see that in the limit $G \to \infty$ our double eigenvalue
problem (\ref{finalab}) is solvable in closed form at any $N = 0,
1, 2,\ldots$.

\subsection*{A. Pascal-like triangle for the Taylor coefficients}

Taylor coefficients ${u}^{(0)}_{[n+1],k}(N), \ k = 0, 1, \ldots,
N$ define the exact wave functions $\psi^{(0)}_{[n]}(r)$
(\ref{anhari}) at each physical root (\ref{fyzi}). Via a suitable
$N-$dependent normalization one may calculate these coefficients
in integer arithmetics (i.e., exactly). This is illustrated in
Table~\ref{vecreals}. At the maximal real roots $s=s_{[1]}(N)=N$
we have
 \ben u^{(0)}_{[1],n}(N)=(-1)^{N+n} \left ( \ba N\\n \ea
\right )
 \een
and our recurrences degenerate to the Pascal triangle for binomial
coefficients,
 \be \left ( \ba N\\n
\ea  \right )
=
\left ( \ba N-1\\n \ea \right ) + \left ( \ba N-1\\n-1 \ea  \right
). \label{Pascalor}
 \ee
This implies the elementary form $\psi(r) =\psi_{[1]}(r) \sim
(1-r/\mu)^N$ of the related exact wave functions. One detects the
presence of a degenerate nodal zero of multiplicity $N$ at
$r=r_z=\mu$. A transition to the next, smaller real root
$s=s_{[2]}(N)=N-3$ makes the multiplicity of the node in $\psi(r)
= \psi_{[2]}(r)$ lowered by two.  The phenomenon survives
iterations in the bracketed subscripts.  Our closed formula
(\ref{fyzi}) for the roots may be complemented by the similar rule
$ N-2n$ for the nodal multiplicities.  With the odd integer
$N=2J+1$ the iterations end at the real spectral root
$s_{[J+1]}(2J+1)=1-J$. A simple, non-degenerate zero appears in
the related wave function $\psi^{(0)}_{[J+1]}(r)$.  This
characterizes the first excitation.  At even $N = 2K$ the final
choice of the minimal $s_{[K+1]}(2K)=-K$ produces the nodeless
wave function $\psi^{(0)}_{[K+1]}(r)$. It describes the ground
state.

A closer inspection of the numerical values of the Taylor
coefficients reveals the presence of certain non-binomial cases.
Most quickly this ``anomaly" is spotted in the ground-state
coefficients ${u}^{(0)}_{[K+1],k}(2K), \ k = 0, 1,\ldots, 2K$. The
same pattern re-appears in the first excitations
${u}^{(0)}_{[J+1],j}(2J+1),\ j= 0, 1,\ldots,2J+1 $ at the higher
$J$ for odd $N=2J+1$, etc.  Thus, the ground-state set is most
fundamental and a few more values ${u}^{(0)}_{[K+1],k}(2K)$ are
displayed in Table~\ref{Pascal}.  It definitely confirms a
surprise.  Each of the coefficients proves to be a {\em sum of its
{\bf three} closest upper neighbors},
 \ben
{u}^{(0)}_{[K+1],k}(2K)= {u}^{(0)}_{[K],k}(2K-2)+
{u}^{(0)}_{[K],k-1}(2K-2)+ {u}^{(0)}_{[K],k-2}(2K-2).
 \een
In an unexpected parallel of the above two-term rule
(\ref{Pascalor}) our Table~\ref{Pascal} forms a new Pascal-like
triangle. For ground states with different $K$ we have
$\psi^{(0)}_{[1]}(r) = 1$, $\psi^{(0)}_{[2]}(r) = 1 + r/\mu +
r^2/\mu^2$, $\psi^{(0)}_{[3]}(r) = (1 + r/\mu+ r^2/\mu^2)^2$ and
so on. We may generalize this result also to all the wave
functions $\psi^{(0)}_{[n+1]}(r)$ which correspond to the same
(even or odd) integer $N= 0, 1, \ldots$,
\be
\psi^{(0)}_{[n+1]}(r) = (1-r/\mu)^{N-2n}\,(1 + r/\mu +
r^2/\mu^2)^n, \ \ \ \ \ \ \ \ \ n = 0, 1, \ldots, \left [
\frac{N}{2} \right ].
 \label{hore}
 \ee
This is one of our main results valid, by induction, for all the
bracketed subscripts.

\subsection*{B. Roots of the coupled secular equations}

The multiple, seemingly degenerate nodal zeros of solutions
(\ref{hore}) require a ``magnification" of their vicinity by an
appropriate inclusion of a few higher-order corrections ${\cal
O}(1/\Omega)$, ${\cal O}(1/\Omega^2 ),\,\ldots$. The only
exceptions are the particular minimal-root ground states
 \ben
 \psi^{(0)}_{[K+1]}(r) =
  (1 + r/\mu + r^2/\mu^2)^K,\ \ \ \ \ N = 2K,
 \ \ \ \ \ \ s = t = -K,
 \een
and the first excitations
 \ben
 \psi^{(0)}_{[J+1]}(r) = (1-r/\mu)\,(1
  + r/\mu + r^2/\mu^2)^J, \ \ \ \ \ N =
 2J+1,
  \ \ \ \ \ \ s = t = 1-J
 \een
without multiple zeros. In the strong-core phenomenological regime
they may prove useful for immediate applications in principle. In
such a context even the most simplified schematic example $V(r) =
A\,r^4+G/r^2$ with $G \gg A$ indicates that the forces with large
$G \gg 1$ need not necessarily lose a reasonable physical
interpretation. A minimum of this function lies at a point $r_0=
(G/2A)^{1/6}$ which only slowly moves with $G$. Whenever
necessary, this motion may further be slowed down by a
simultaneous increase of the coupling $A$.

In order to get a feeling for the subtleties of structure of the
systems with large $G$ let us pick up our ground-state solution
with $N=4$ and $s=t=-K=-2$ in $s-$wave ($\ell=0$). For simplicity
let us assume the absence of the Coulombic and cubic terms in our
potential $V_{\{2\}}(r)$. This means $F=B=0$ and fixes the values
of $\beta = 0$ and $\gamma=C/2\alpha$ with $C=4(A^2 /\Omega)^{
1/3}$. At the ground-state energy $ E = -\gamma^2-4 (A \Omega)^{
1/3}$ our special example
 \be
 V_{\{2\}}(r) = \alpha^2r^4 +4(\alpha^4/\Omega)^{1/3}r^2
 -2\alpha(\Omega+5)\,r+{\Omega(\Omega-1)}/{r^2}
 \label{test}
 \ee
is still only solvable in the limit $\Omega \to \infty$.
Nevertheless, due to our explicit knowledge of the wave function,
we may now invert the procedure easily. All the higher-order
deviations from solvability may already be added to eq.
(\ref{test}) in an explicit form produced, say, via the insertion
of our elementary ground state formula for $\psi_{[2]}(r)$ in the
original, differential Schr\"{o}dinger eq. (\ref{rad})).

For definitness of our illustration let us choose the unit
spring-constant coefficient $C=1$ at $r^2$ in $V_{\{2\}}(r)$. We
get, as a consequence, a strong quartic confinement $\alpha^2=A =
\sqrt{\Omega}/8$ combined with the repulsive linear force
possessing an even stronger, ${\cal O}(\Omega^{5/4})$ coupling $D=
-\Omega^{1/4}(\Omega+5)/\sqrt{2}$. We may re-scale the coordinate
$r \to r \Omega^{1/4}$ in our radial differential Schr\"{o}dinger
equation (\ref{rad}) with force (\ref{test}). In the leading-order
$1+{\cal O}(1/\Omega)$ approximation this leads to the zero-energy
problem with the dominant three-term interaction $ V_{\{2\}}(r) =
const\,r^4 -const'\,r +const''/r^2$. All its coupling constants
are of the same order of magnitude $ {\cal O}(1)$. It is known
\cite{rnad} that the solutions of such a problem may also be
elementary and proportional to Laguerre polynomials.

We may conclude that in the limit $\Omega \to \infty$ the whole
picture and scheme of elementary solvability is nicely
self-consistent. We also see the usefulness of our present
generalized construction. It leads to a richer class of polynomial
solutions $\psi(r)$ in a fairly nontrivial though still feasible
and manageable way. What remains for us is to extend the same
pattern of self-consistency to all the higher-order (i.e.,
presumably, $1/\Omega^k$) corrections.

\section*{IV. Corrections
\label{III}}

At the finite values of the coupling $G$ the strength of the
perturbation $\lambda\,Q^{(1)}= Q - Q^{(0)} $ is measured by the
parameter $\lambda=1/(\mu \tau)= 1/\Omega$. Its smallness is
strictly equivalent to our above assumption $\Omega \gg N$. In a
perturbative treatment of our bound-state problem $Q\,\vec{u}=0$
we have to avoid the following two most serious obstacles.

Firstly, our pseudo-Hamiltonian $Q(s,t)$ of eq. (\ref{scaled}) and
its unperturbed simplification $ Q^{(0)}(s,t)$ as well as their
three-diagonal difference
 \ben
Q^{(1)}=\left(
\begin{array}{cccccc}
0&0&&&&  \\ 0&-\gamma\mu&1 &&&
\\ &-\beta\mu^2&-2\gamma\mu&3 &&
\\& &-2\beta\mu^2&-3\gamma\mu&\ddots &
\\
 & &&-3\beta\mu^2&\ddots &
{N(N-1) \over 2}
 \\
 & & && \ddots&
-N\gamma \mu\\
 & & & & & -N\beta \mu^2\\
\end{array} \right)
 \een
are non-square matrices. This means non-linearity, not tractable
by the standard perturbation formalisms. Secondly, our zero-order
pseudo-Hamiltonian itself is a four-diagonal matrix.  This would
cause difficulties even in the linear case, with an unclear idea
how one could construct an unperturbed propagator. Fortunately,
the apparently unavoidable methodical pessimism is not in place.
With certain care one can proceed in an almost complete analogy
with the traditional Rayleigh-Schr\"{o}dinger textbook
prescription,

\begin{itemize}

\item
choosing an arbitrary integer $N\geq 0$,

\item
picking up a real root $s^{(0)}=t^{(0)}$ as given by eq.
(\ref{fyzi}),

\item
postulating, in nonstandard manner, the {\em two} different
expansions
 \ben
  s=s^{(0)}+\lambda\, s^{(1)}+\lambda^2 s^{(2)}+\ldots
 \een
 \ben
  t=t^{(0)}+\lambda\, t^{(1)}+\lambda^2 t^{(2)}+\ldots \een
plus, more traditionally,
 \ben \left( \ba u_0\\ u_1\\  \vdots\\
u_N \ea \right)= \left( \ba u^{(0)}_0\\ u^{(0)}_1\\  \vdots\\
u^{(0)}_N \ea \right) +\lambda\,\left( \ba u^{(1)}_0\\
u^{(1)}_1\\  \vdots\\ u^{(1)}_N \ea \right) +\lambda^2 \left(
\ba u^{(2)}_0\\ u^{(2)}_1\\  \vdots\\ u^{(2)}_N
\ea \right)+\ldots,
 \een

\item
combining all these expansions with input $ Q=
Q^{(0)}+\lambda\,Q^{(1)} $.

\end{itemize}

 \noindent
As a net result we get the hierarchy of equations for corrections.
On all the subsequent levels of precision ${\cal O}(\lambda^k)$,
i.e., in the $k-$th perturbation order with $k=1, 2, \ldots$,
these equations will play the role of implicit definitions of the
``charge", ``energy" and ``wave function" corrections $s^{(k)}$,
$t^{(k)}$ and $\vec{u}^{(k)}$, respectively. The set will be
initiated by the above zero-order problem (\ref{finale}). All the
subsequent new equations will have the same non-square and
non-homogeneous common matrix form
 \ben
 \left [ Q^{(0)}(0,0) +s^{(0)}\,{\cal J}+t^{(0)}\,{\cal K}
 \right ]
 \vec{u}^{(k)}+
\left [ Q^{(1)}(0,0) +s^{(1)}\,{\cal J}+t^{(1)}\,{\cal K}
 \right ]
 \vec{u}^{(k-1)}+
 \een
 \be
 +
\left ( s^{(2)}\,{\cal J}+t^{(2)}\,{\cal K}
 \right )
 \vec{u}^{(k-2)}+ \ldots +
\left ( s^{(k)}\,{\cal J}+t^{(k)}\,{\cal K}
 \right )
 \vec{u}^{(0)}=0.
 \label{finauj}
 \ee
{\it Mutatis mutandis}, we may parallel the standard textbook
perturbation theory of the Rayleigh-Schr\"{o}dinger type
\cite{Messiah}.

\subsection*{A. Perturbations of the charges and energies}

Our first important observation is that eq. (\ref{finale}) and
properties of our zero-order pseudo-Hamiltonians
$Q^{(0)}(s^{(0)},t^{(0)})= Q^{(0)}(0,0)+s^{(0)}\,{\cal J} +
t^{(0)}\,{\cal K}$ imply that these non-square and $(N+2)\times
(N+1)-$dimensional matrices always possess the {\em pairs of
independent left eigenvectors}. Their explicit sample
$[\vec{v}^{[s]}]^T$ and $[\vec{v}^{[a]} ]^T$ is displayed in the
first part of Table~\ref{leftvecs}. We normalized these vectors of
dimension $N+2$ in the symmetric and antisymmetric manner,
respectively. Wherever necessary, the Dirac's symbols will be used
to denote the similar ``longer" columns as ``kets" (e.g.,
$\vec{v}^{[s] }\equiv | {v}^{[s]}\rangle$) and the rows as ``bras"
(say, $[\vec{v}^{[s]}]^T \equiv\langle {v}^{[s]}|$ etc).

Marginally, let us admit that there exist alternative
possibilities of normalization. The requirement of disappearance
of the first or last matrix element in a suitable linear
superposition of $ \langle v^{[s/a]}|$ may be used to define the
``compactified" left eigenvectors $\langle w_\star^{}
|\equiv\vec{\sigma}^T{\cal J}^T$ and $\langle w^{\star }
|\equiv\vec{\theta}^T{\cal K}^T$. The study of their properties
discouraged us from their use in computations since the new
doublets do not always remain linearly independent.  This is
exemplified at $N=2$ and $s^{(0)}=2$ in the second part of
Table~\ref{leftvecs}.

The left action of a left auxiliary eigenbra $ \langle
v^{[s/a]}|$ completely eliminates the vector $ \vec{u}^{(k)}$
from eq.  (\ref{finauj}).  We abbreviate
\ben
 |\Xi^{(k-1)} \rangle = -\left [ Q^{(1)}(0,0) +s^{(1)}\,{\cal
J}+t^{(1)}\,{\cal K} \right ] \vec{u}^{(k-1)}- \een
\ben
 -
\left ( s^{(2)}\,{\cal J}+t^{(2)}\,{\cal K}
 \right ) \vec{u}^{(k-2)}- \ldots -
\left ( s^{(k-1)}\,{\cal J}+t^{(k-1)}\,{\cal K}
 \right ) \vec{u}^{(1)}
\een and arrive at the two independent equations \ben \langle
v^{[s]}|
 ({\cal J} \vec{u}^{(0)}) \rangle
 \,s^{(k)} + \langle
v^{[s]}|
 ({\cal K} \vec{u}^{(0)}) \rangle \, t^{(k)}= \langle
v^{[s]}|\Xi^{(k-1)} \rangle ,
 \een
  \ben \langle v^{[a]}|( {\cal J}
\vec{u}^{(0)}) \rangle \,s^{(k)} + \langle v^{[a]}| ({\cal K}
\vec{u}^{(0)}) \rangle \, t^{(k)}=
 \langle v^{[a]}|\Xi^{(k-1)} \rangle .
 \een
They may be interpreted as a two-by-two matrix inversion. The
symmetric and antisymmetric normalization of our auxiliary
eigenvectors implies that $\langle v^{[s]}| ({\cal
J}\vec{u}^{(0)}) \rangle =\langle v^{[s]}| ({\cal K }\vec{u}^{(0)}
)\rangle$ and $\langle v^{[a]}| ({\cal J}\vec{u}^{(0)}) \rangle
=-\langle v^{[a]}| ({\cal K }\vec{u}^{(0)} )\rangle$.  This
enables us to invert the left-hand-side matrix in an explicit
manner,
 \ben \left ( \ba s^{(k)}
\\ t^{(k)} \ea \right ) = \frac{1}{2}\, \left (
\begin{array}{cc}
1/\langle v^{[s]}|
({\cal J} \vec{u}^{(0)}) \rangle&
1/\langle v^{[a]}|
({\cal J} \vec{u}^{(0)} )\rangle\\
1/\langle v^{[s]}|
({\cal K} \vec{u}^{(0)} )\rangle&
1/\langle v^{[a]}| ({\cal K} \vec{u}^{(0)} )\rangle
\ea
\right )
\left (
\ba
 \langle v^{[s]}|\Xi^{(k-1)} \rangle
\\
 \langle v^{[a]}|\Xi^{(k-1)} \rangle
\ea \right ).
 \een
This is our first final closed formula for corrections $ s^{(k)}$
and $t^{(k)}$, an extended non-linear parallel to the usual
Rayleigh-Schr\"{o}dinger definition of energies.

\subsection*{B. Perturbations of the wave functions}

The column vector of corrections $\vec{u}^{(k)}$ is characterized
by its re-normalization ambiguity $\vec{u}^{(k) }\to \vec{u}^{(k)}
+ const \times \vec{u}^{(0)} $. This follows from the very
definition (\ref{finauj}) and implies that, say, the first
component of $\vec{u}^{(k)}$ may be chosen as vanishing. Let us
indicate such an option by a superscript $^\star$ in
$\vec{u}^{(k)} =\vec{u}^{\star (k)}$ with ${u}_{0}^{\star (k)}=0$
for $k > 0$.  In our definition (\ref{finauj}) this choice of
normalization makes the pseudo-Hamiltonian matrix $
Q^{(0)}(s^{(0)},t^{(0)})$, in effect, lower triangular.  In the
same equation all the known terms may be collected in the
single ket \ben |\tau^{(k-1)}\rangle= |\Xi^{(k-1)} \rangle
-s^{(k)}\,{\cal J}\vec{u}^{(0)} -t^{(k)}\,{\cal K}\vec{u}^{(0)}
\een
entering the right-hadn side of our set of relations $
Q^{(0)}(s^{(0)},t^{(0)} )\,\vec{u}^{\star (k)}
=|\tau^{(k-1)}\rangle$. Two of its rows (or, more precisely, the
two linear combinations of all these rows) have already been
used for the determination of the quantities $s^{(k)}$ and
$t^{(k)}$.  We are left with the $N$ independent equations. In the
second decisive step of our method we recommend their choice
which drops the last two lines and offers the relations
\ben R^\star\,
\left ( \ba u_1^{\star (k)}\\ u_2^{\star (k)}\\ \vdots \\
u_N^{\star (k)} \ea \right ) =  \left ( \ba \tau_0^{\star (k-1)}\\
\tau_1^{ \star (k-1)}\\ \vdots
\\ \tau_{N-1}^{ \star (k-1)} \ea \right )
  \een
with
 \ben R^\star= \left(
\begin{array}{cccccc}
1&&&&&\\ s^{(0)}&2&&& & \\ t^{(0)}&s^{(0)}&3 && & \\
N-1&t^{(0)}&s^{(0)}&4 &&
\\ &\ddots&\ddots&\ddots &\ddots& \\ & &3 & t^{(0)}&s^{(0)} &N
\end{array} \right)
 \een
i.e., an $N-$dimensional linear matrix equation which defines
all the remaining unknown quantities.  Computationally, it just
offers the highly economical evaluation of the higher-order
corrections $\vec{u}^{(k)}$ via the downwards-running and finite
four-term recurrences.

An important merit of our proposal is that the main diagonal of
our lower triangular (and finite-dimensional) matrix $R^\star$ is
safely non-zero. Hence, the matrix itself is regular, $\det
R^\star = N! \neq 0$. Its lower triangular inversion $\left
(R^\star \right)^{-1}$ exists and may be interpreted as the
present analogue of the current Rayleigh-Schr\"{o}dinger
unperturbed propagator.

The direction of our recurrences is an inessential consequence of
the choice of normalization. After its alternative specification
$\vec{u}^{(k)} =\vec{u}_\star^{(k)}$ with $({u}^{(k)
}_\star)_{N}=0$ in all the orders $k > 0$ we annihilate the last
column of $Q^{(0)}(s^{(0)},t^{(0)})$. Naturally, one omits now the
first two rows. Thus, we cut the $N-$dimensional and upper
triangular submatrix $R_\star$ out of~$Q^{(0)}(s^{(0)},t^{(0)})$
and define the alternative unperturbed propagator by inversion,
 \ben \left ( \ba u_{\star\,1}^{(k)}\\ u_{\star\,2}^{(k)}\\
\vdots \\ u_{\star\,N}^{(k)} \ea \right ) = \left ( R_\star \right
)^{-1}\, \left ( \ba \tau_{\star\,0}^{(k-1)}\\
\tau_{\star\,1}^{(k-1)}\\ \vdots
\\ \tau_{\star\,N-1}^{(k-1)} \ea \right ).
  \een
Computationally, of course, the higher-order corrections
$\vec{u}_\star^{(k)}$ are again much more easily evaluated
directly, via the upwards-running four-term recurrences.


\section*{V. Conclusions \label{IV}}

The core of our present paper may be seen in the progress we
achieved in the non-numerical Magyari-like constructions. By means
of their decisive perturbative simplification we were able to
offer new solutions of certain ``next-to-solvable" models.

In the first step we analyzed the limit $G \to \infty$  of quartic
forces $V_{\{2\}}(r)$ and were able to find an exact and complete
solution of the underlying nonlinear algebraic equations. This
demonstrates that far beyond the ``safe" territory of the exact
and quasi-exact solutions one may still get closed formulas much
more easily than expected before.

Another pleasant surprise appeared during our study of corrections
${\cal O}(1/\Omega)$. For many years we felt deterred by the
strong $N-$dependence of their traditional numerical analyses as
well as by the quick increase of complexity of their explicit
description with $m\geq 2$ and dimensions $N$ beyond, say, $N=0$
of $N=1$ \cite{Flessas}. In this context, the simple form of our
present quasi-linear and $N-$independent perturbative formalism
opens a new direction of development towards many practical (say,
phenomenological) applications yet to be constructed and
appreciated.

We have demonstrated that the traditional Taylor-series approach
\cite{Ince} to the differential Schr\"{o}dinger equation may still
offer new surprises. Its main merit lies in its ability of
combination of the correct threshold behaviour with the correct
physical asymptotics of $\psi (r) $ for a broad class of
elementary potentials. As a consequence, many simple models in
quantum mechanics may efficiently be clarified by means of the
related ``most natural" power series ansatzs:

\begin{itemize}
\item
In place of the differential equation one gets a (sometimes more
transparent) set of recurrences;
\item
Their unique solution may often be written in a closed (often
called Hill-determinant \cite{Classif}) form at any trial energy
$E$;
\item
The truncated power-series wave functions remain fairly precise in
a finite (i.e., most relevant) interval of coordinates even after
a rough energy guess.

\end{itemize}

\noindent

The transparency of the recipe is often marred by its various
practical disadvantages. For example, the efficiency of the
Hill-determinant algorithms often lags behind the universal
variational approaches \cite{Hills}. Also the domains of their
applicability may be much more constrained, the unpleasant
phenomenon which leads to many non-trivial misunderstandings
\cite{Chaudhuri}. Difficulties emerge even in numerical
implementations of these algorithms: In the fixed-precision
computer arithmetics one has to avoid the potential instability of
recurrences etc \cite{Tater}. In such a context the most reliable
remedy lies, obviously, in an {\em exact} termination of the
infinite series.

In our paper we emphasized that the disadvantages connected with
the latter strategy are often overestimated in applications. We
succeeded in showing that the merits of using the algebraic
Magyari-like equations may often be preserved at a reasonable
price. Proceeding in a consequently constructive way we have shown
that their non-linearity has, in a way, a ``weak" form which
admits a certain ``quasi-linear" treatment.

This attitude is of course trivial for the models of the Coulomb,
Kratzer or harmonic oscillator type. For all of them the power
series method is able to generate {\em all} their exact
bound-state solutions as terminating. Their exact solvability
involves all the angular momenta $\ell$ and physical energies $E$.
Similarly, under weaker assumptions, the same polynomial ansatzs
also succeed in offering the ``quasi-exact" \cite{Ushveridze}
multiplets of bound states of a finite size~$N$.

The {\em freedom} of choosing the multiplicity $N$ has been
significantly extended here. Having noticed that the latter two
``solvable" categories mean mostly the two- and three-term
character of the underlying recurrences, we paid thorough
attention to their very next, four-term generalization. In the
context of an enormous interest in quartic oscillators
\cite{dalsi} we imagined that one need not always perceive the
comparatively short four-term recurrences as prohibitively
complicated. At the same time, for many phenomenological
applications, they offer a partially non-numerical insight into
the new class of interactions which is already fairly rich.

In the light of our present results, the nonlinearity of the
underlying algebra need not represent a serious technical
obstacle. This is our main message. Via a detailed analysis of our
present quartic polynomial examples we have shown that the
explicit solution of the related Schr"{o}dinger and Magyari
equations remains feasible in an almost complete parallel with
their quadratic predecessors.

\newpage

\begin{table}
\caption{The key to our nonlinear problem (\ref{finalab}):
Its roots $s^{(0)} =t^{(0)}$  for
the first few $N\neq 0$. } \label{list}
\begin{center}
\begin{tabular}{||c||rrrrrrrrrrrr||}
 \hline \hline   && &&&&&&&&&&\\ $N$ &
1& &2&&&&3&&&&&\\   && &&&&&&&&&&\\ \hline
 2 Re $s^{(0)}$ &
2&-1&4&1&-2&-2&6&3&0&0&-3&-3\\ 2 Im $s^{(0)}/\sqrt{3}$& 0&$\pm$
1&0&$\pm$ 1& 0& $\pm$ 2& 0& $\pm$ 1& 0& $\pm$ 2& $\pm$ 1& $\pm$
3\\ \hline  \hline   && &&&&&&&&&&\\ $N$ & 4&& &&&&&&&&&\\   &&
&&&&&&&&&&\\ \hline
 2 Re $s^{(0)}$ &
8&5&2&2&-1&-1&-4&-4&-4&&&\\
 2 Im $s^{(0)}/\sqrt{3}$&
0&$\pm$ 1& 0& $\pm$ 2& $\pm$ 1& $\pm$ 3 &
 0& $\pm$ 2& $\pm$ 4&&&\\
 \hline  \hline  & && &&&&&&&&&\\ $N$ & 5&& &&&&&&&&&\\  & && &&&&&&&&&\\
  \hline
 2 Re $s^{(0)}$ &
 10&7&4&4&1&1&-2&-2&-2&-5&-5&-5\\
 2 Im $s^{(0)}/\sqrt{3}$&
0&$\pm$ 1& 0& $\pm$ 2& $\pm$ 1& $\pm$ 3 &
 0& $\pm$ 2& $\pm$ 4&
 $\pm$ 1& $\pm$ 3 &
  $\pm$ 5
 \\
 \hline \hline
\end{tabular}
\end{center}
\end{table}

\newpage

\begin{table}
\caption{ Real eigenvectors of our pseudo-Hamiltonians $Q^{(0)}$.}
\label{vecreals}
\begin{center}
\begin{tabular}{||cr|rrrrrrr||}
\hline \hline
 &&&&&&&&\\
  $N$ &$s^{(0)}$ & $u^{(0)}_0$ & $u^{(0)}_1$ & $u^{(0)}_2$
& $u^{(0)}_3$ & $u^{(0)}_4$ & $u^{(0)}_5$ & $u^{(0)}_6$
 \\  &&&&&&&&\\
 \hline
 \hline
   0
& 0 & 1&-&-&-&- &-&-\\
 \hline 1 & 1 & 1&-1&-&-&- &- &-\\
  \hline 2
& 2 & 1&-2&1&-&-  &-&-\\
 & -1 & 1&1&1&-&- &- &-\\
\hline 3 & 3 & 1&-3&3&-1&-  &-&-\\
 & 0 & 1&0&0&-1&- &- &-\\
\hline 4 & 4 & 1&-4&6&-4&1  &-&-\\
 & 1 & 1&-1&0&-1&1 &- &-\\
 & -2 & 1&2&3&2&1  &-&-\\
\hline 5 & 5 & 1&-5&10&-10&5  &-1&-\\
 & 2 & 1&-2&1&-1&2  &-1&-\\
 & -1 & 1&1&1&-1&-1  &-1&-\\
\hline 6 & 6 & 1&-6&15&-20&15  &-6&1\\
 & 3 & 1&-3&3&-2&3  &-3&1\\
 & 0 & 1&0&0&-2&0  &0&1\\
 & -3 & 1&3&6&7&6  &3&1\\
\hline \hline
\end{tabular}
\end{center}
\end{table}

\newpage

\begin{table}
\caption{ Generalized Pascal triangle (ground states: $N=2K$,
$s_{[K+1]}=-K$). } \label{Pascal}
\begin{center}
\begin{tabular}{||c|ccccccccccc||}
\hline \hline
 &&&&&&&&&&&\\
 $K$ & \ldots& $u^{(0)}_{K-4}$ & $u^{(0)}_{K-3}$ &
$u^{(0)}_{K-2}$ & $u^{(0)}_{K-1}$ & $u^{(0)}_{K}$ &
$u^{(0)}_{K+1}$ & $u^{(0)}_{K+2}$ & $u^{(0)}_{K+3}$ &
$u^{(0)}_{K+4}$&\ldots
 \\  &&&&&&&&&&&\\ \hline  &&&&&&&&&&&\\
 0 & \ldots& 0 &
0&0& 0& 1&0& 0&0& 0& \ldots\\ 1 & \ldots& 0 & 0&0& 1& 1&1& 0&0& 0&
\ldots\\ 2 & \ldots& 0 &0& 1& 2& 3&2& 1& 0& 0&\ldots\\ 3 & \ldots&
0&1 & 3& 6& 7&6& 3& 1& 0&\ldots\\ 4 & \ldots& 1&4 & 10& 16& 19&16&
10& 4&1& \ldots\\ 5 & \ldots& 5&15 & 30& 45& 51&45& 30& 15&
5&\ldots\\ 6 & \ldots& 21& 50 & 90& 126& 141&126& 90& 50&
21&\ldots\\ \ldots &&&&&&\ldots&&&&&\\ \hline \hline
\end{tabular}
\end{center}
\end{table}

\begin{table}
\caption{ Pairs of independent left eigenvectors of
$Q^{(0)}(s,s)$.} \label{leftvecs}
\begin{center}
\begin{tabular}{||crr||rrrrrr||rrrrr||c||}
\hline \hline &&&
 \multicolumn{6}{c||}{sym. and antisym. $v_k$}&
 \multicolumn{5}{c||}{shortened $\sigma_k$ and $\theta_k$} & overlaps \\
 \hline
 \hline
 && $k=$&0&1&2&3&4&5&0&1&2&3&4&\\
  $N$ &$s$ &&
&&&&&&&&&&& $\langle v |({\cal J} \vec{u}^{(0)})\rangle$
\\
 \hline
 \hline
   0& 0& & 1&1&-&-&- &-& 1&-&-&-&-&1\\
   &  & &1&-1&-&-&- &-& 1&-&-&-&-&1\\
 \hline
 1 & 1 && 1&-2&1&-&- &- & 1&-1&-&-&-&3\\
   &  & &1&0&-1&-&- &- & - 1&1&-&-&-&1\\
  \hline
    2& 2 && 2&-1&-1&2&-  &-& 0&1&-1&-&-&3\\
  &  & &0&1&-1&0&-  &-& - &-&-&-&-&-3\\
   & -1 && 1&1&1&1&- &- & 3&1&2&-&-&3\\
 &  & &3&-1&1&-3&- &- & 2&1&3&-&-&3\\
\hline
 3 & 3 && 1&1&-2&1&1  &-& 1&0&-1&1&-&-9\\
  &  & &1&-1
  &0&1&-1  &-& 1&-1&0&1&-&3\\
 & 0 & &2&-1&0&-1&2 &- & 2&0&0&-1&-&3\\
 &  & &2&1&0&-1&-2 &- & - 1&0&0&2&-&3\\
\hline
 4 & 4 && 1&-2&1&1&-2  &1& 1&-1&0&1&-1&9\\
  &  & &1&0&-1&1&0  &-1& - 1&1&0&-1&1&-9\\
   & 1& & 7&1&-2&-2&1 &7 & 7&-3&-1&-1&4&9\\
 &  & &1&-1&0&0&1 &-1 & 4&-1&-1&-3&7&3\\
 & -2 && 1&1&1&1&1  &1& 3&1&2&1&2&9\\
 &  & &3&-1&1&-1&1  &-3& 2&1&2&1&3&3\\
\hline \hline
\end{tabular}
\end{center}
\end{table}

\end{document}